\documentclass[prd,reprint,onecolumn,notitlepage,preprintnumbers, showpacs,superscriptaddress]{revtex4-1}
\usepackage{verbatim}
\usepackage{amsmath, multirow, amssymb, wasysym, gensymb}
\usepackage{graphicx}
\usepackage{units}
\usepackage{color}
\usepackage[utf8]{inputenc}
\begin{document}
\title{Knotted strings and leptonic flavor structure}
\author{T. W. Kephart}
\affiliation{Department of Physics and Astronomy, Vanderbilt University, Nashville, TN 37235, USA}
\author{P. Leser}
\email{philipp.leser@tu-dortmund.de}
\affiliation{Fakult\"at f\"ur Physik,
Technische Universit\"at Dortmund, 44221 Dortmund,
Germany}
\author{H. P\"as}
\email{heinrich.paes@tu-dortmund.de}
\affiliation{Fakult\"at f\"ur Physik,
Technische Universit\"at Dortmund, 44221 Dortmund,
Germany}
\preprint{DO-TH 11/18}
\pacs{14.60.Pq, 14.60.St, 02.10.Kn}

\begin{abstract}
We propose a third idea for the explanation of the leptonic flavor structure in addition to the prominent approaches based on flavor symmetry and anarchy. Typical flavor patterns can be modeled by using mass spectra obtained from the discrete lengths spectrum of tight knots and links.
We assume that a string theory model exists in which this idea can be incorporated via the Majorana mass structure of a type I seesaw model. It is shown by a scan over the parameter space that such a model is able to provide an excellent fit to current neutrino data and that it predicts a normal neutrino mass hierarchy as well as a small mixing angle $\theta_{13}$. Startlingly, such scenarios could be related to the dimensionality of space time via an anthropic argument.
\end{abstract}
\maketitle
\section{Introduction}
One of the most profound problems in contemporary particle physics is the 
flavor puzzle or the question whether the observed flavor structure is governed
by anarchy (essentially random numbers) 
\,\cite{Hall:1999sn,Hirsch:2001mw,deGouvea:2003xe,Altarelli:2002sg}
or a (typically discrete) flavor symmetry (see e.g., Refs.~\cite{Ma:2004zd,Ma:2007ia,Altarelli:2010gt,Ishimori:2010au}).
In this paper, we demonstrate that there exist further alternatives. More precisely, we will propose that the leptonic flavor structure could arise
from the topological configurations of closed strings.

Closed strings are a fundamental
ingredient of string theory, including in particular
the graviton and its AdS/QCD dual, the glueball, as well as dilaton
superfields with fermionic degrees of freedom having the correct
quantum numbers of a right-handed neutrino (a fact used extensively e.g., in neutrino
mass models with large extra dimensions, see e.g.,~Ref.~\cite{ArkaniHamed:1998vp}).

It thus seems well-motivated that topologically nontrivial string or flux tube configurations such as knots and
links can contribute to the mass of closed string states, and may even dominate it. As string tension tends to minimize the string length, the
knot or link length can be assumed to be directly proportional to the mass.
For example, it has been shown in Refs.~\cite{Buniy:2002yx,Buniy:2004zy,Buniy:2010cu}, that
the experimental spectrum of glueball candidates can be fitted very nicely
by knot and link energies. An application of a knot model to flavor physics has recently been discussed in Ref.~\cite{Finkelstein:2010bp}.

Here we exploit another interesting feature of the knot and link spectrum. Typically there exist different close-to-degenerate states with very small energy gaps.
If right-handed neutrino masses are dominated by the knots and links
of closed strings in a seesaw framework, large and maximal leptonic mixing
may result naturally from the knot and link spectrum without the need
for any flavor symmetry.

\section{Description of the Model and Predictions}
 Now consider a generic type I seesaw model, in which the flavor structure originates from
 the mass matrix of the right-handed sector, which is 
generated by a large scale mass spectrum of knotted strings or flux tubes. 
Consequently the Dirac masses are assumed to be diagonal:
\begin{align}
	m_\text{D} &= 
	\begin{pmatrix}
		m_1^\text{D} & 0   & 0\\
		0   & m_2^\text{D} & 0\\
		0   & 0   & m_3^\text{D}
	\end{pmatrix},
\end{align}
while the symmetric Majorana mass matrix has the following structure:
\begin{align}
	M &= 
	\begin{pmatrix}
		m_1^\text{K} & m_1^\text{L} & m_2^\text{L}\\
		m_1^\text{L} & m_2^\text{K} & m_3^\text{L}\\
		m_2^\text{L} & m_3^\text{L} & m_3^\text{K}
	\end{pmatrix}.
\end{align}

This way the choice of the Dirac masses is mainly responsible for the absolute mass scale of
neutrinos, while the choice of the structure in the right-handed sector 
implies the leptonic mixing matrix. 
This is the general structure of mass matrices analyzed in Ref.~\cite{Leser:2011fz}.

We assume that the heavy masses $m_i^{\text{K}/\text{L}}$ take on values according to the spectrum of characteristic lengths of knots and links\,\cite{Ashton:2010fk}, multiplied by a common scale $m_\text{S}$ at which knotted string configurations can exist:
\begin{align}
	m_i^{\text{K}/\text{L}} = \ell_i^{\text{K}/\text{L}}\cdot m_{\text{S}},
\end{align}
where the $\ell_i^{\text{K}/\text{L}}$ refer to the characteristic lengths of Table~\ref{tab:lengths}. The diagonal entries of the mass matrix are generated by the knots' lengths, while the off-diagonal entries are related to the characteristic lengths of the links.

The mass matrix of the three light left-handed neutrinos in the flavor basis is then obtained using the usual formula $M_\text{flv}^\nu=m_\text{D}^T M^{-1} m_\text{D}$:
\begin{align}
	M_\text{flv}^\nu&=\frac{1}{\Delta^3}
	\begin{pmatrix}
	 \left[m_2^\text{K} m_3^\text{K}-\left(m_3^\text{L}\right)^2\right] \left(m_1^\text{D}\right)^2
	    & (m_2^\text{L} m_3^\text{L}-m_3^\text{K} m_1^\text{L}) m_1^\text{D} m_2^\text{D} & (m_1^\text{L} m_3^\text{L}-m_2^\text{K} m_2^\text{L})
	   m_1^\text{D} m_3^\text{D} \\
	 (m_2^\text{L} m_3^\text{L}-m_3^\text{K} m_1^\text{L}) m_1^\text{D} m_2^\text{D}
	    & \left[m_1^\text{K} m_3^\text{K}-\left(m_2^\text{L}\right)^2\right] \left(m_2^\text{D}\right)^2 & (m_1^\text{L} m_2^\text{L}-m_1^\text{K} m_3^\text{L})
	   m_2^\text{D} m_3^\text{D} \\
	 (m_1^\text{L} m_3^\text{L}-m_2^\text{K} m_2^\text{L}) m_1^\text{D} m_3^\text{D}
	    & (m_1^\text{L} m_2^\text{L}-m_1^\text{K} m_3^\text{L}) m_2^\text{D} m_3^\text{D} & \left[m_1^\text{K} m_2^\text{K}-\left(m_1^\text{L}\right)^2\right]
	   \left(m_3^\text{D}\right)^2
	\end{pmatrix},
	\label{eqn:flavormatrix}
\end{align}
where the common factor of mass dimension three is given by $\Delta^3 = -m_3^\text{K} \left(m_1^\text{L}\right)^2+2m_{1}^\text{L} m_{2}^\text{L} m_{3}^\text{L}-m_2^\text{K} \left(m_2^\text{L}\right)^2-m_1^\text{K} \left(m_3^\text{L}\right)^2+m_{1}^\text{K}m_{2}^\text{K}m_{3}^\text{K}$. The $(1,1)$ element of Eq.~\eqref{eqn:flavormatrix} is the effective mass $m_{\beta\beta}$ observed in neutrinoless double beta decays.

It is instructive to examine the structure of the effective mass matrix analytically to determine some general features of the model. As the tribimaximal mixing pattern\,\cite{Harrison:2002er} still provides a reasonable approximation for the experimental data, we look at the general structure needed to generate such mixing angles. In general, a mass matrix that leads to tribimaximal mixing can be parametrized as\,\cite{Harrison:2003aw}
\begin{align}
M_\nu^\text{TBM} &=	\begin{pmatrix}
		x & y & y\\
		y & x+v & y-v\\
		y & y-v & x+v
	\end{pmatrix},
	\label{eq:genmatrix}
\end{align}
where $x$, $y$ and $v$ are real numbers.

Assuming that a normal mass hierarchy can be approximated as two vanishing neutrino masses and one  neutrino mass at a higher scale $\tilde{m}$---i.e. a diagonal mass matrix of $\text{diag}(0,0,\tilde{m})$---this leads to a mass matrix of the form
\begin{align}
	\tilde{m}\cdot\begin{pmatrix}
		0 & 0 & 0\\
		0 & \frac{1}{2} & -\frac{1}{2}\\
		0 & -\frac{1}{2} & \frac{1}{2}
	\end{pmatrix},
\end{align}
which can be thought of as setting $x=0, y=0, v=1/2$ in Eq.~\eqref{eq:genmatrix}. This matrix is then compared to the mass matrix of Eq.~\eqref{eqn:flavormatrix}. The comparison yields a set of relations between the Majorana parameters $m_i^\text{K},m_i^\text{L}$ and the other parameters:
$m^\text{K}_3/m^\text{K}_2 = \left(m_2^\text{D}\right)^2/\left(m_3^\text{D}\right)^2$, $m^\text{L}_2/m^\text{L}_1 = m_3^\text{D}/m_2^\text{D}$, $m^\text{K}_2 m^\text{L}_2 = m^\text{L}_1 m^\text{L}_3$, $m^\text{K}_1 m^\text{K}_2 \neq \left(m^\text{L}_1\right)^2$, $\tilde{m} = 2 \left(m_3^\text{D}\right)^2 \left(\left(m^\text{L}_1\right)^2 - m^\text{K}_1m^\text{K}_2\right)/\Delta^3$.
If the Dirac masses $m_i^\text{D}$ are assumed to be roughly equal the first three conditions can be fulfilled if the selected lengths $\ell^\text{K}_i$ and $\ell^\text{L}_i$ are close to each other. In general, as the order (crossing number) of the knots increases, the spacing decreases since the length grows roughly linearly with crossing number, but the number of knots grows faster than exponentially with crossing number.

As the neutrino mass scale $\tilde{m}$ is small due to the seesaw mechanism, the electroweak scale $m_3^\text{D}$ factor in the condition $\tilde{m} = 2 \left(m_3^\text{D}\right)^2 \left(\left(m^\text{L}_1\right)^2 - m^\text{K}_1m^\text{K}_2\right)/\Delta^3$ needs to be compensated by making the expression in the parentheses small; this can again be achieved by having an almost degenerate spectrum for $\ell^\text{K}_i$ and $\ell^\text{L}_i$. Since the spectrum of knots and links features almost degenerate lengths, it 
is thus expected that it will provide a better fit
to the leptonic flavor structure than random numbers.

The corresponding condition for an inverted hierarchy, which is approximated as two neutrino masses at a higher scale $\tilde{m}$ and one neutrino mass set to zero---i.e. the diagonal mass matrix $\text{diag}(\tilde{m},\tilde{m},0)$, leads to a mass matrix
\begin{align}
	\tilde{m}\cdot\begin{pmatrix}
		1 & 0 & 0\\
		0 & \frac{1}{2} & \frac{1}{2}\\
		0 & \frac{1}{2} & \frac{1}{2}		
	\end{pmatrix}.
\end{align}
Comparing this to Eq.~\eqref{eqn:flavormatrix} gives a system of equations that can only be solved if $\tilde{m}=0$. Thus, in this approximation it is not possible to generate an inverted neutrino mass hierarchy. Taking into account that the tribimaximal pattern is only an approximation and that the smaller mass difference is not zero, one would expect that in this model the inverted mass hierarchy should be suppressed.

Finally  we analyze the compatibility of the model with a degenerate neutrino mass spectrum. Assuming a diagonal mass matrix $\text{diag}(\tilde{m},\tilde{m},\tilde{m})$ the conditions that follow from  Eq.~\eqref{eqn:flavormatrix} read: $m_1^\text{D} m_3^\text{D} \neq 0$, $m^\text{K}_2 = m^\text{K}_1 \left(m_2^\text{D}\right)^2/\left(m_1^\text{D}\right)^2$, $m^\text{K}_3 = m^\text{K}_1 \left(m_3^\text{D}\right)^2/\left(m_1^\text{D}\right)^2$, $m^\text{K}_3 \neq 0$, $m^\text{K}_1m^\text{K}_2 \neq 0$, $m^\text{K}_1 \neq 0$, $m^\text{L}_1 =m^\text{L}_2 = m^\text{L}_3 = 0$, $m^\text{K}_1m^\text{K}_2\left(m_3^\text{D}\right)^2 + \tilde{m}\cdot\Delta^3 = 0$.

Out of these conditions, the last two are in contradiction with the framework of the model: The $m^\text{L}_i$ and $m^\text{K}_i$ parameters cannot be zero or close to zero. The model investigated in this paper thus cannot be used to explain a degenerate neutrino mass hierarchy.

\section{Numerical Analysis}
In order to investigate the viability of the models, every possible combination of characteristic lengths up to a given knot order is sampled using a computer code. No duplicate lengths of knots or links are allowed.

The parameters $m_i^\text{D}$ for $i=1,\ldots, 3$ as well as the overall scale of the Majorana masses are not fixed by the model. As the scope of this analysis is the viability of the choice of knots and links as a source of Majorana masses, the Dirac masses are chosen in a way as to minimize the $\chi^2$ value of the squared mass differences of the neutrinos compared to experimental data\,\cite{GonzalezGarcia:2010er}. This way, no potentially viable combinations of knots and links are discarded due to a wrong choice for the Dirac masses. The overall Majorana scale factor that is multiplied with the characteristic lengths of the knots and links is fixed at $10^{12}$\,GeV.

If the characteristic string spectrum is realized by cosmic strings, one has to respect bounds obtained by the effect of such cosmological defects on the power law index of primordial density perturbations as
measured in CMB probes such as WMAP\,\cite{Urrestilla:2011gr}.

Such cosmological defects arise in the phase transitions associated with the spontaneous 
breakdown of non-Abelian gauge symmetries.
The string tension, which is Newton's constant times the mass
per unit length, is then related to the symmetry breaking scale. If strings are formed at the GUT scale
$10^{16}$\,GeV, then the string tension is approximately $10^{-6}$, which is below the constraint
from CMB observations\,\cite{Urrestilla:2011gr}.
Even stronger constraints result from the contribution of cosmic strings to the 
stochastic background of gravitational waves which can be constrained from pulsar timing
observations\,\cite{Olmez:2010bi}. These constraints
require a string tension below $10^{-9}$ corresponding to a symmetry breaking scale of 
about $10^{13}$\,GeV. Consequently we adopt this value as an upper bound for the scale confinement
$m_\text{conf}=m_\text{S}$.

For the subset of models that have acceptable squared mass differences the mixing angles are calculated and also compared to the experimental values. All models with a $\chi^2<16.8$ for the mixing angles are considered viable. This corresponds to a $P$ value of $0.01$ and six degrees of freedom.

The scan covering 10,692,864 possible combinations of knots' and links' lengths results in 321,781 models with normal neutrino mass hierarchy and 8731 models with an inverted neutrino mass hierarchy that fall below the $\chi^2$ limit of $16.8$. This means that about $3.1\%$ of all possible combinations yield phenomenologically acceptable results. The best fit lies in the regime of normal hierarchy with a $\chi^2_\text{best} = 0.001$. The best fit model is described by the parameters in Table~\ref{tab:bestfitmodel}.

This has to be compared with a model based on random numbers: The same model structure is assumed, but the list of characteristic lengths is replaced by a list of random numbers between $0$ and the largest knot length of the actual list of knot lengths. This is repeated for 10 sets of random numbers. The total number of acceptable models is $1.6\%$ of all tested combinations. In all cases, the best fit model is in the normal hierarchy regime and the total number of viable models with a normal neutrino mass hierarchy is much larger than the number of models with an inverted mass hierarchy. There are two effects that explain this discrepancy: In our models with an inverted mass hierarchy, $\theta_{13}$ is usually predicted to be close to maximal, while the models with normal mass hierarchy predict a naturally small $\theta_{13}$.

In addition to this, the relative number of models with a normal neutrino mass hierarchy is even larger in the case of knots and links. This can be explained by the conditions that follow from  Eq.~\eqref{eqn:flavormatrix}, which lead to the spectrum of knots and links being able to fit the requirements for a normal mass hierarchy easier than random numbers.

For all random cases considered the total number of acceptable models is lower than the total number of acceptable models in the case of knots and links. This means that the models using the characteristic lengths of knots and links are more suitable to fit the neutrino data than a fit using random numbers.

\begin{table}[h]
\centering
\begin{tabular}{c|c|cccccc|ccc|c}
Hierarchy & $\chi^2$ &$K_1$ & $K_2$ & $K_3$ & $L_1$ & $L_2$ & $L_3$ & $m^\text{D}_1$\,[GeV] & $m^\text{D}_2$\,[GeV] & $m^\text{D}_3$\,[GeV] & Scale factor [GeV]\\
\hline
Normal   & 0.001 & 01 & 06 & 11 & 11 & 17 & 12 & $12.193$ & $13.207$ & $12.867$ & $1\cdot 10^{12}$\\
Inverted & 0.09 & 02 & 08 & 10 & 08 & 00 & 17 & $59.601$ & $16.441$ & $14.986$ & $1 \cdot 10^{12}$ 
\end{tabular}
\caption{The model parameters giving the best fit for normal and inverted hierarchies. The knots and links indices refer to table \ref{tab:lengths}.}
\label{tab:bestfitmodel}
\end{table}

To determine the phenomenological consequences of the allowed models, the following observables are calculated: the double beta decay parameter $m_{\beta\beta}$, the lightest neutrino mass $m_0$ and the neutrino mixing angle $\theta_{13}$. In the normal hierarchy case, $m_{\beta\beta}$ tends to be small, i.e. between $0.001$\,eV and $0.01$\,eV. Note that the best fits yield values for $m_{\beta\beta}$ between $0.001$\,eV and $0.007$\,eV. In the case of an inverted mass hierarchy, $m_{\beta\beta}$ takes on values between $0.01$\,eV and $0.02$\,eV.

As the angle $\theta_{13}$ is small and the contribution from $m_0$ is negligible, this is in line with the results from Ref.~\cite{Leser:2011fz}, where the parameter $m_{\beta\beta}$ is given as:
\begin{align}
	m_{\beta\beta} &\approx
	\begin{cases}
		\sqrt{\Delta m^2_{12}}\sin^2\left(\theta_{12}\right) & \text{for normal hierarchy}\\
		\sqrt{\Delta m^2_{\text{23}}}\quad\text{resp.}\quad\sqrt{\Delta m^2_{\text{23}}} \cos\left(2\theta_{12}\right) & \text{for inverted hierarchy} 
	\end{cases}.
\end{align}

\begin{figure}[p]
	\centering
		\includegraphics[width=5.5in]{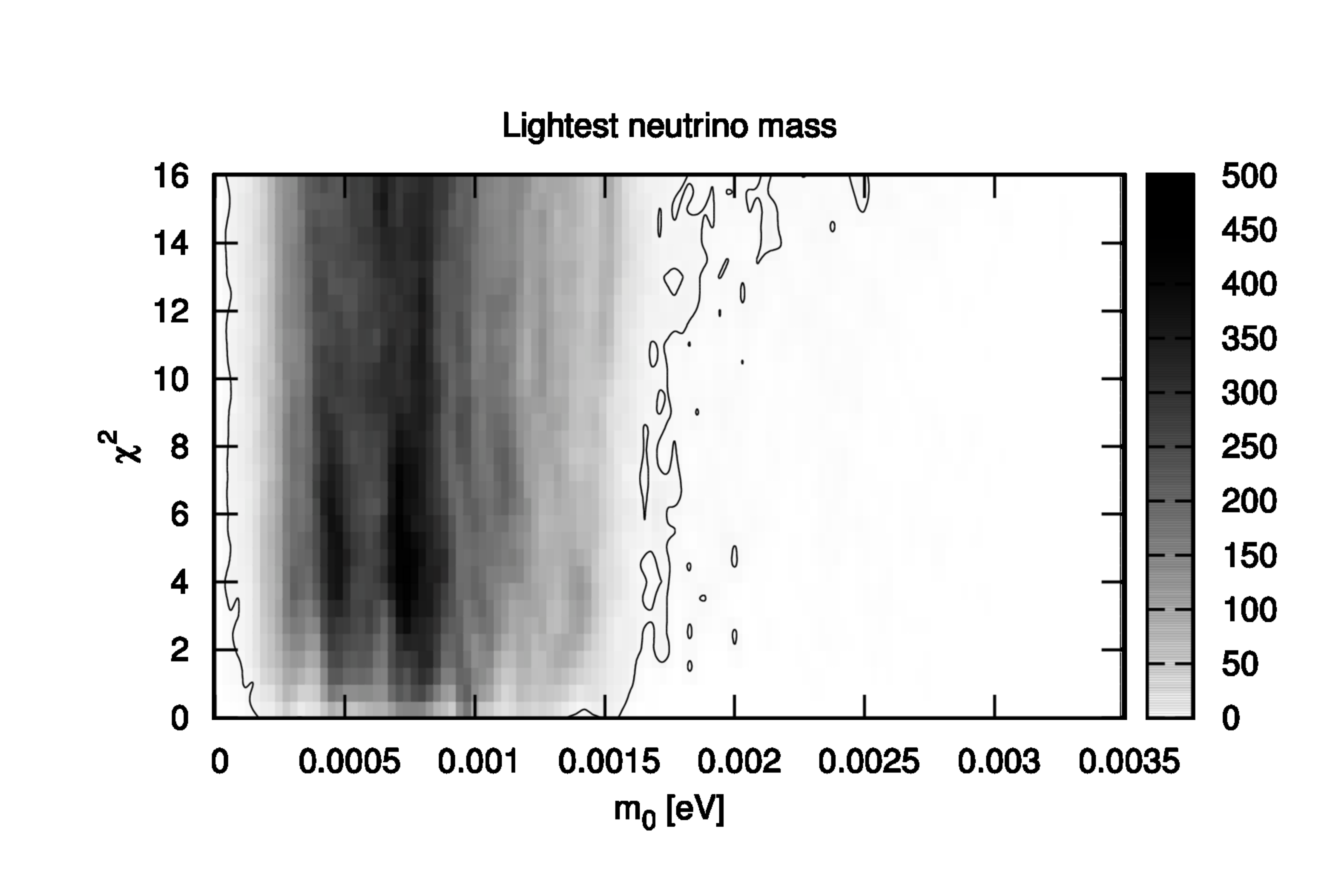}
	\caption{The lightest neutrino mass $m_0$ for the models with a normal neutrino mass hierarchy. The plot has been divided into $121$ bins along the $x$-axis and $35$ bins along the $y$-axis. The shade of the rectangles represents the number of models found in that area. Outside of the boundary line, less than ten hits per rectangle were recorded.}
	\label{fig:klnormlightest}
\end{figure}

The lightest neutrino mass in the normal hierarchy case is below $0.003$\,eV, illustrated in Fig.~\ref{fig:klnormlightest}. For an inverted mass hierarchy, a lightest mass of up to $0.007$\,eV is possible. In both cases, the neutrino masses are well below the current bound on the sum of the neutrino masses $\sum m_i \lesssim 0.5$\,eV\,\cite{Steidl:2009hx,Elgaroy:2002bi,Tegmark:2006az,Hannestad:2006mi,Spergel:2006hy,Seljak:2006bg,Allen:2003pta,   Hannestad:2006as,Hannestad:2007cp,Lesgourgues:2004ps}.

\begin{figure}[p]
		\centering
			\includegraphics[width=5.5in]{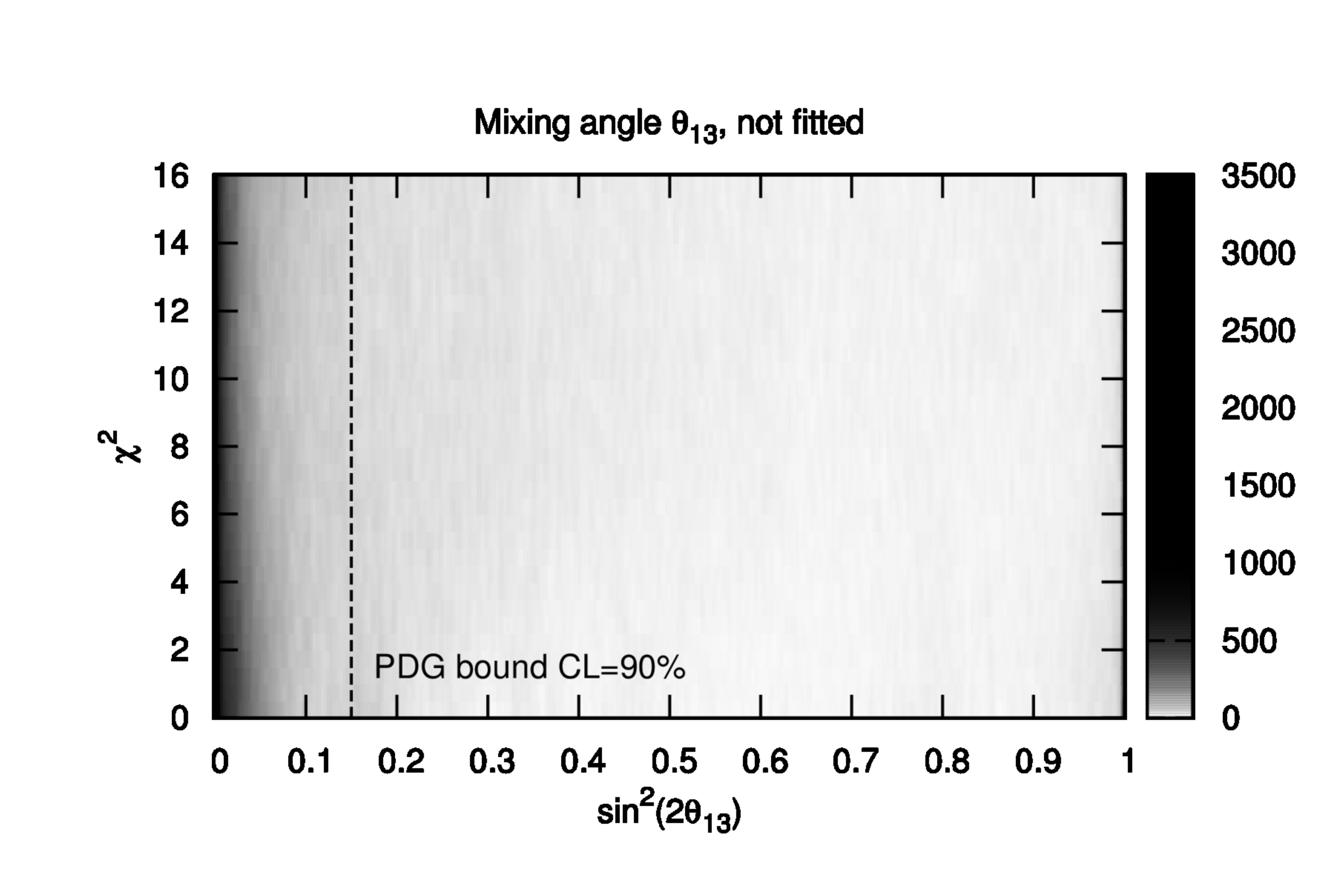}
		\caption{The quantity $\sin^2\left(2\theta_{13}\right)$ for the models with a normal mass hierarchy. $\theta_{13}$ was excluded from the fit for this plot. The plot has been divided into $334$ bins along the $x$-axis and $35$ bins along the $y$-axis. The shade of the rectangles represents the number of models found in that area. The bound on $\theta_{13}$ given by Ref.~\cite{Nakamura:2010zzi} is indicated as a dashed line.}
		\label{fig:not13norm}
	\end{figure}

Using the framework of a global fit of experimental data\,\cite{GonzalezGarcia:2010er} we have fitted the model to the squared masses and mixing angles, but omitted $\theta_{13}$ from the calculation. We have then surveyed the model prediction for the angle and compared that to the global fit, but also to the recent experimental results of T2K\,\cite{Abe:2011sj} and Daya Bay\,\cite{An:2012eh}. This has been done for normal and inverted neutrino mass hierarchies and the results for a normal mass hierarchy are shown in Fig.~\ref{fig:not13norm}. In the case of a normal neutrino mass hierarchy, a small angle $\theta_{13}$ close to zero is preferred. In the case of inverted mass hierarchy, a very large angle $\theta_{13}$ is preferred, although some results are still below the applicable bound. For a comparison with possible values for $\theta_{13}$ generated by models based on discrete symmetries, see Ref.~\cite{Albright:2006cw}.

Taking the results from the T2K experiment\,\cite{Abe:2011sj} into account a range for $\sin^2 2\theta_{13}$ is given as $0.03(0.04) < \sin^2 2\theta_{13} < 0.28(0.34)$ for a normal (inverted) neutrino mass hierarchy at 90\% CL. Within the allowed $\chi^2$ range assumed here, the results of this paper are compatible with these bounds.

Our results are therefore also compatible with the current best fit value of $\sin^2 2\theta_{13} = 0.092$ given by the Daya Bay collaboration in Ref.~\cite{An:2012eh}.

\section{Possible UV Completions}
We would like to stress that the scenario we are pursuing here is an effective model which may result from various ultraviolet completions. In the following we sketch some qualitative ideas about such completions.
First, fundamental closed strings could be considered, but it is not clear if these can have tight knots because of their vanishingly small cross section.
Another option to generate massive knots near the GUT scale are cosmic strings. If a collapsing loop 
of nontrivial topology $K$ tightens before it decays, then the tight knot  configuration will have
mass
 \begin{align}
M_\text{K} \sim L_\text{K} \langle \phi \rangle,
\end{align} 
near the symmetry
breaking scale $\langle \phi \rangle$, where the $U(1)$ is broken that gives rise to the cosmic string.
Here $L_\text{K}$ is the dimensionless length of the knot $K$, i.e. the length of the knot divided by the radius of the cosmic string.

If a knot is bosonic  above the SUSY breaking scale, then it will also have a fermionic partner of the same mass.
Furthermore, 
If the fermionic knots are gauge singlets, then they can serve as the heavy right-handed
neutrinos needed for the  seesaw
mechanism to generate the very light observed neutrino states.

The stability of various knot types will be model dependent, hence the lightest knots may not be stable and so may not be
the ones that mix with the light neutrinos.
To arrive at yet a different possibility to generate the spectrum of knots and links we use in this work, consider a ten dimensional $E_8\otimes E'_8$ heterotic superstring theory  and compactify it on a
Calabi--Yau manifold $K$ with $SU(3)$ holonomy. With a proper choice of $K$
and identification of its holonomy group with a subgroup of $E_8$ we can arrive
at a four dimensional $E_6\otimes E_8'$ theory with three chiral $E_6$ families\,\cite{Candelas:1985en}, i.e.  three 27s of $E_6$.
 The three family $SU(3)\times SU(2)\times U(1)$ standard model is embedded in the $E_6$ sector and has been studied. But we will focus our attention
on the $E'_8$ hidden sector, which has only gravitational interactions with standard model 
particles.

Let the $E_6$ and $E_8'$ gauge couplings be unified a string scale  $M_\text{string} \sim 5 \times10^{17}$ GeV.  Then we expect
$\alpha_\text{string}(M_\text{string}) \sim 1/50$ in order that the SM couplings agree with experiment. Since
the hidden sector has no chiral fermions, $\alpha_{E'_8} $ runs quickly to $O(1) $ at the mass scale 
approaches $M_{8'} \sim 10^{13}$ GeV.

The hidden sector has been assumed to generate supersymmetry breaking via a gluino
condensation when the theory becomes non-perturbative\,\cite{Derendinger:1985kk,Dine:1985rz} at $\alpha_{E'_8}(M_{8'}).$ In addition,  the $E'_8$ theory becomes confining near this energy scale, but since there are no fundamental chiral fermions in the hidden sector, there are
no ``light'' ($\sim 10^{13}$ GeV) mesonic or baryonic $E'_8$ states. The lightest particles in the hidden sector  will be closed   
$E'_8$ flux loops. These solitons will be glueball like and we expect their spectrum to scale like the tight knot/link spectrum\,\cite{Buniy:2002yx,Buniy:2004zy}. Furthermore, supersymmetry will either be unbroken, or an approximate symmetry at the $M_{8'}$ scale, so these solitons will have 
equal mass fermionic partners. We expect bosonic glueballs to have $J^{PC}=0^{++}$ quantum numbers, and their fermionic partners to
be $\frac{1}{2}^{++}$ states with no standard model quantum numbers. 
The lightest $E'_8$ glueball is nearly stable (but can decay by quantum reconnection) because there are no lighter hidden sector states into which it can decay. The entire $E'_8$ knot spectrum is metastable since each knot-soliton carries a different topological charge derivable from its individual unique knot invariants.
 
This analysis leads us to suggest that the knotted fermionic hidden sector solitons
provide a natural source of neutral heavy singlet fermions for the seesaw mechanism in order  to give mass to light neutrinos. Our observation that the low energy neutrino data is better fit by a  knot spectrum than by a random mass spectrum 
may indicate hidden sector dynamics, i.e. hidden sector confinement and tightly knotted flux tube formation that provide a viable model for the leptonic flavor structure.

Finally, the scenario presented here features a curious anthropic twist which may relate it to the dimensionality of spacetime: If we assume that the baryon asymmetry is generated via leptogenesis from the decay of the heavy right-handed neutrinos, it becomes important that knotted strings are stable only in spacetimes with three spatial dimensions. In universes with a larger number of dimensions the knots would untie and the states corresponding to the heavy Majorana neutrino would not exist. Consequently no baryons would be generated and no intelligent observers could evolve. According to the spirit of anthropic reasoning the consequence that intelligent observers would be possible only in those landscape vacua which feature three space dimensions could be understood as an argument for the observed dimensionality of space time.

\section{Conclusions}
In summary, we have proposed an alternative to both flavor anarchy and flavor symmetry:
a seesaw type I model whose Majorana mass structure is governed by the discrete spectrum of tight knots and links. A possible UV completion of this model may result from
supersymmetric cosmic strings arising from the GUT scale breaking of an additional $U(1)$ gauge
symmetry, or from fermionic partners of glueball-like states originating as hidden sector flux
tubes   
 in ten-dimensional $E_8\otimes E'_8$ heterotic superstring theory. Based on the general structure of the mass matrices of Ref.~\cite{Leser:2011fz}, we have shown that the model fits the current experimental neutrino data on squared mass differences and mixing angles. It has also been shown that the spectrum of knots and links produces a larger number of viable models than a spectrum of random numbers. The model favors a normal neutrino mass hierarchy and predicts a small mixing angle $\theta_{13}$. Startlingly, such scenarios could be related to the dimensionality of space time via an anthropic argument.

\section*{Acknowledgments}
This work was supported by DFG grant PA 803/6-1. The work of TWK was supported by DOE grant number DE-FG05-85ER40226. PL was supported by the Studienstiftung des deutschen Volkes. 

%

\begin{table}[h]
\centering
\begin{tabular}{c|c|c|c|c|c|c|c|c|c|c}
	Index 				&	00		&	01		&	02		&	03		&	04		&	05		&	06		&	07		&	08		&	09\\ \hline
	Knot length [a.u.]	&	32.7436
	                    &42.0887
	                    &47.2016
	                    &49.4701
	                    &56.7058
	                    &57.0235
	                    &57.8392
	                    &61.4067
	                    &63.8556
	                    &63.9285\\	                   	
	Link length [a.u.]	&25.1334
	                    &40.0122
	                    &49.7716
	                    &54.3768
	                    &56.7000
	                    &58.1013
	                    &57.8141
	                    &58.0070
	                    &50.5539
	                    &64.2345\\ \hline\hline
	Index 				&	10	&	11	&	12	&	13	&	14	&	15	&	16	&	17&&\\ \hline
						Knot length [a.u.]  &64.2687
						                    &65.2560
						                    &65.6924
						                    &65.6086&&&&&\\		
						Link length [a.u.]	&65.0204
						                    &65.3257
						                    &65.0602
						                    &66.1915
						                    &66.3147
						                    &55.5095
						                    &57.7631
						                    &65.8062&&
\end{tabular}
\caption{Characteristic lengths of knots and links up to knot order 7, taken from Ref.~\cite{Ashton:2010fk}}
\label{tab:lengths}
\end{table}
\end{document}